\documentstyle[12pt]{article}
\begin{document}
\baselineskip 1.2\baselineskip
\pagestyle{empty}
\begin{flushright}
Edinburgh-/91-92/01 KAIST-THP-92/702 YITP/U-92-19
\end{flushright}
\begin{center}
{\Large {\bf Singularity analysis in $A_n$ Affine Toda Theories}}
\end{center}
\begin{center}
{\bf{H.W. Braden}} \\
Department of Mathematics, University of Edinburgh, \\
Edinburgh, UK. \\
{\bf{H.S. Cho, J.D. Kim, I.G. Koh}}         \\
Physics Department     \\
Korea Advanced Institute of Science and Technology        \\
373-1 Kusongdong Yusonggu, Taejon 305-701, Korea.        \\
{\bf{R. Sasaki}} \\
Uji Research Center, \\
Yukawa Institute for Theoretical Physics, \\
Kyoto University, Uji 611, Japan.   \\
\end{center}
\begin{center}
ABSTRACT
\end{center}
The leading and the subleading Landau singularities in affine
Toda field theories are examined in some detail.
Formulae describing the subleading simple pole structure of
box diagrams are given explicitly. This leads to a new and nontrivial test
of the conjectured exact S-matrices for these theories.
We show that to the one-loop level the conjectured S-matrices of the  $A_n$
Toda family reproduce the correct singularity structure,
leading as well as subleading, of the field theoretical amplitudes.
The present test has the merit of being independent of the details of the
renormalisations.

\newpage
\pagestyle{plain}

\section*{I. Introduction}

The affine Toda  field theories (ATFTs) are a remarkable class of massive
two-dimensional  models. Classically  integrable,  these models
possess candidate exact S-matrices conjectured to describe the
quantum theory\cite{AFZ,BCDSc,SMat,DVeF,DGZs}. Such
S-matrices enable one to study the thermodynamic
Bethe ansatz\cite{ZKM} and other scaling properties of the theory. Although
these S-matrices have passed many nontrivial low order checks,
a proof has yet to be given that they are indeed the S-matrices
of the theory.
These models remain an attractive testing ground for understanding
some of the rich and diverse phenomena of quantum field theory.
This letter will study some of complicated Landau singularity\cite{ELOP}
structure of ATFT and provide a new and nontrivial test
for the putative exact S-matrices.

The coupling constant $\beta$ dependence of the exact ATFT S-matrices is
believed to appear through a single universal function $B(\beta)$
\cite{AFZ,BCDSc,SMat},
\begin{equation}
B(\beta) = \frac{1}{2 \pi} \frac{\beta^2}{1+ \beta^2/4\pi},
\label{BBETA}
\end{equation}
and this has been verified to $\beta^4$ by conventional perturbation
theory in the absence of the anomalous threshold singularities\cite{DGZs,BS}.
Recently the $\beta^6$ term was  also confirmed in $A_2$ ATFT using a
dispersion relation approach\cite{SZ}.
This example is particularly simple for the
triangle diagrams that appear are nonsingular and one only needs
to consider the effect of renormalisation and not the effect of
the (here vanishing) Landau singularity\cite{BSa}.
In general at this order delicate cancellations appear between
the leading Landau singularities of the  relevant box and triangle diagrams
\cite{BCDSe}
as well as the  subleading terms\cite{CKKR,CKK}.
This letter will
further test the conjectured S-matrices by examining
the role of these subleading  terms when they too are singular.
We will focus on $A_n$ Toda theories for two reasons.
First, the present singularity analysis up to double poles is complete for the
$A_n$ series.
Second, the effects of renormalisation can be clearly separated
from those of the Landau singularity in these cases.
We show how the subleading
Landau singularities conspire to give the correct
(in fact vanishing) residues of the simple poles at the general
double pole positions of the $A_n$ theory's S-matrices.

The general methods of extracting subleading singularities developed here are
universally applicable to any two-dimensional field theories.
However, the complete singularity analysis for the other members of ATFTs,
$D_n$ and $E_n$ series, is further complicated by the higher order poles
and renormalisation effects.

\section*{II. Preliminaries}
We will now state our conventions and recall the essential
points of ATFT needed for our calculation.
The bosonic ATFT\cite{MOP,BCDSc} based upon a Lie algebra $g$ has rank $r$
massive scalar fields $\phi^{a}$ with exponential interactions.
The Lagrangian of the theory is \footnote{For various reasons \cite{BCDSc}
we restrict to ATFTs based on simply laced algebras.}
\begin{equation}
{\cal{L}}(\phi) = \frac{1}{2}\partial_{\mu}\phi^{a}\partial^{\mu}\phi^{a}
-\frac{m^{2}}{\beta^{2}}\sum_{i=0}^{r}n_{i}e^{\beta \alpha_{i}^{a} \cdot
\phi^{a}},
\end{equation}
where $\alpha_{i}$ ($i=1,\cdots,r$) are the simple roots of $g$
(normalised $\alpha_i\sp2 = 2$) and
\begin{displaymath}
\alpha_{0} = -\sum_{i=1}^{r}n_{i}\alpha_{i},~~ ~~ n_{0} = 1 .
\end{displaymath}
The integers $n_i$ are the so called Kac labels
for the Lie algebra.
Here $m$ sets the mass scale and the real number $\beta$ is the dimensionless
coupling constant.

These theories may be described equally well by listing the masses and
multipoint coupling constants. Classically the ${\rm mass}\sp2$
are given by the eigenvalues of the matrix
\begin{equation}
(M^{2})^{ab} = m^{2} \sum_{i=0}^{r}n_{i}\alpha_{i}^{a}\alpha_{i}^{b}
\end{equation}
and the three-point couplings may be obtained from
\begin{equation}
c^{abc} = m^{2}\beta \sum_{i=0}^{r} n_{i}\alpha_{i}^{a}\alpha_{i}^{b}
\alpha_{i}^{c}.
\end{equation}
The four-point and higher point couplings are determined by this
information\cite{BSa}.
For  the $A_n$ theories the classical masses are
\begin{equation}
m_a = 2m\sin\frac{a\pi}{h} , ~~~~a=1,\dots,n,
\end{equation}
(where $h=n+1$ is the Coxeter number for $A_n^{(1)}$)
and the nonzero multipoint couplings are neatly given  in terms of these by
\begin{equation}
C_{a_1\dots a_p} =(-1)^{\frac{\sum{a_i}}{h}}(-i)^p
(\frac{\beta^2}{m^2h})^{\frac{p}{2}-1}
\prod_{k=1}^{p}m_{a_k}, ~~~~~ {\rm if}~~ \sum_{k=1}^{p}a_k \equiv 0 ~~
({\rm mod}~h).
\label{VER}
\end{equation}

The two-particle elastic S-matrices for the $A_n^{(1)}$ theories are
conjectured to be
\begin{equation}
S_{ab} = \prod_{|a-b|+1 ~ {\rm step} 2}^{a+b-1}\{p\},
\label{SMAT}
\end{equation}
where, using the notation of ref.\cite{BCDSc}\footnote{For an interesting
compact representation of the S-matrices in terms of vertex operators see
\cite{CDor}.},
\begin{equation}
\{x\}=\frac{(x-1)(x+1)}{(x-1+B)(x+1-B)},~~~ (x)=\frac{\sinh(\frac{\theta}{2}
+\frac{i\pi x}{2h})}{\sinh(\frac{\theta}{2}-\frac{i\pi x}{2h})},
\label{BBLOCK}
\end{equation}
and $B(\beta)$ is given by Eq.(\ref{BBETA}).
These S-matrices have maximally double poles.

Finally we  summarise the relevant Feynman rules  adopted
\begin{equation}
\begin{array}{ll}
(a)~ \mbox{For each (three point) vertex} &:  -i(2\pi)^2 c^{abc},  \\
(b)~ \mbox{For each propagator}   &: \frac{i}{(2\pi)^2(p_a^2-m_a^2)},   \\
(c)~ \mbox{Loop integration}      &: \int d^2 l .
\end{array}
\label{FRULE}
\end{equation}
In order to obtain the S-matrix element from the Feynman amplitude
we need the flux normalisation factor or Jacobian coming from the
change of variables from the linear momentum to the rapidity \cite{BSa}.
With $p_a=m_a(\cosh\theta_a,\sinh\theta_a)$ and $\theta =\theta_a-\theta_b$
this behaves near the singular point $\theta =i\theta_0$ as
\begin{equation}
\frac{1}{(2\pi)^2 4m_am_b\sinh\theta}\nonumber
=\frac{-i}{(2\pi)^2 4m_am_b\sin\theta_0}[1+i\cot\theta_0(\theta-i\theta_0)+
\cdots].
\label{FLUX}
\end{equation}
For future reference we also note the following relationship between
the Mandelstam  variable $s=(p_a+p_b)^2$ and rapidity $\theta$  near
the singularity
\begin{eqnarray}
\frac{1}{s-s_0} =
\frac{-i}{2m_am_b\sin\theta_0(\theta-i\theta_0)}
[1+\frac{i\cot\theta_0}{2}(\theta-i\theta_0)+ \cdots].
\label{MAND}
\end{eqnarray}

\section*{III. The Subleading Singularity}
A general Feynman amplitude
may be expanded  about some point $i\theta_0$ in the rapidity as follows
\begin{equation}
\mbox{Amplitude }=\frac{R_{-p}}{(\theta-i\theta_0)^p}+\frac{R_{-p+1}}
{(\theta-i\theta_0)^{p-1}}+\cdots +R_0+R_1(\theta-i\theta_0)+ \cdots.
\end{equation}
Here $p$ is the maximal order of the singularity and
the tool most often used in the analysis of these singular terms
is the so called \lq scaling' method\cite{BCDSe,DdeVa,GPZa,CKKR,CKK}.
We shall now describe how this works for the leading and
subleading terms of the uncrossed
box diagram (a) and the crossed box diagrams (b) and (c) of figure 1.

If we denote the $i$-th internal propagator momentum as $Q_i$,
the momentum integral for the general box diagram is
\begin{eqnarray}
L(s) & =& \int d^2k \frac{1}{[Q_A^2-m_A^2][Q_B^2-m_B^2][Q_C^2-m_C^2]
[Q_D^2-m_D^2]},   \\
\smallskip
  Q_i &=& q_i+k~=~q_i+(s-s_0)l,\qquad i=A,B,C,D.  \nonumber
\end{eqnarray}
When this diagram has a leading Landau singularity (that is, when
all the internal
propagators become on-shell simultaneously) one finds\cite{BCDSe} that
in the vicinity of the singular point $s_0$
\begin{eqnarray}
L(s) &=& \frac{1}{(s-s_0)^2}\int d^2l \prod_{i}
\frac{1}{[\epsilon_i+2q_i\cdot l +(s-s_0)l^2]}.
\label{scaling}
\end{eqnarray}
If we denote the singular configuration of $Q_i$ by $q_i^{(0)}$ and
the singular configuration of the external particles' momenta
by $p_a^{(0)}$ and $p_b^{(0)}$ then
$\epsilon_i$ here is defined to be the product
of the  two constants $a_i$ and $b_i$ such that
\begin{equation}
q_i^{(0)} = a_i p_a^{(0)}+b_i p_b^{(0)},~~~~
q_i = a_i p_a+b_i p_b, \quad \epsilon_i =a_ib_i.
\label{qqq}
\end{equation}
The constants $a_i$ and $b_i$ can be computed easily from the dual diagram
being the ratios of the area of triangles in the dual diagram.

We are interested in the simple pole contribution from Eq.(\ref{scaling}).
First we note that Eq.(\ref{scaling}) is readily evaluated\cite{BCDSe}
upon the change of variables
\begin{eqnarray}
u=2q_D\cdot l, & v=2q_A\cdot l.
\end{eqnarray}
The corresponding Jacobian can be evaluated by expressing $q_D$ and $q_A$
in terms of the external momenta using
Eq.(\ref{qqq})
\begin{equation}
J=\frac{\partial(l_0,l_1)}{\partial(u,v)}=
\frac{1}{8i\tilde\Delta(a_Db_A-b_Da_A)}.
\end{equation}
Here $\tilde\Delta$ is  $\frac{1}{2}m_am_b\sin\theta$
and the Jacobian for this change of variable is inversely proportional
to the area of the triangle spanned by $q_D$ and $q_A$.
We now observe that there are two sources to
the simple pole contribution of Eq.(\ref{scaling}).
One comes from the next order term in the expansion of $\tilde\Delta$ in
the Jacobian $J$,
\begin{equation}
\frac{1}{\tilde\Delta}=\frac{1}{\Delta}
(1+i\cot\theta_0(\theta-i\theta_0)+\cdots)
\quad~{\rm with}~~\Delta=\frac{1}{2}m_am_b\sin\theta_0
\end{equation}
and the second arises when we expand the integrand
in terms of $(s -s_0)$,
\begin{equation}
\prod_{i} \frac{1}{[\epsilon_i+2q_i\cdot l +(s-s_0)l^2]}=
\prod_{i} \frac{1}{[\epsilon_i+2q_i\cdot l]}\times\big\lbrace
1-\sum_{j}{(s -s_0)l^2\over{[\epsilon_j+2q_j\cdot l]}}+\ldots\big\rbrace.
\label{int}
\end{equation}

The first of these contributions from the Jacobian is the easiest to
evaluate as we have simply the product of the double pole residue
of the appropriate  box diagram\cite{BCDSe} multiplied by $i\cot\theta_0$.
Thus we have corresponding to the diagrams of figure 1
\begin{eqnarray}
(a) &:& (\frac{\beta}{\sqrt{2h}})^4\frac{1}{(\theta-i\theta_0)}
(\frac{2\Delta_a^{'}}{\Delta})i\cot\theta_0 \times S.
\label{SLFBa} \\
(b) &:& (\frac{\beta}{\sqrt{2h}})^4\frac{1}{(\theta-i\theta_0)}
(\frac{\Delta_b-\Delta_a^{'}}{\Delta})i\cot\theta_0 \times S.
\nonumber  \\
(c) &:& (\frac{\beta}{\sqrt{2h}})^4\frac{1}{(\theta-i\theta_0)}
(\frac{\Delta_a-\Delta_b}{\Delta})i\cot\theta_0 \times S.
\nonumber
\end{eqnarray}
Here $S$  is the symmetry factor that takes into account the possible distinct
diagrams giving the same amplitude.

The contribution coming from expanding the integrand requires a little
more work. The integrals are  evaluated by closing contours
in the lower $(u,v)$ half-planes where, under the condition
$\Delta_a^{'} \geq \Delta_b$
(which avoids the appearance of extra poles), there are
poles at $u=-\epsilon_D-i\varepsilon$ and $v=-\epsilon_A-i\varepsilon$.
Using the results of \cite{BCDSe} such as
$\Delta \equiv \Delta_a+\Delta_a^{'} \equiv
\Delta_b+\Delta_b^{'}$
and the expressions for $q_i$ in terms of the triangles and
on-shell momenta together with the relevant
vertex (\ref{VER}) and flux (\ref{FLUX}) factors we are rewarded with
the following nice formulas for the simple pole residues
of the box diagrams
\begin{eqnarray}
(a): & (\frac{\beta}{\sqrt{2h}})^4
\frac{i\Delta_a^{'}}{2\Delta^2\Delta_b \Delta_b^{'} }[
(p_b \cdot p_b )\Delta_a (\Delta_b-\Delta_b^{'})
+(p_a \cdot p_b )(-2\Delta_b \Delta_b^{'})] \times S.
\label{SLFBb}  \\
(b): & (\frac{\beta}{\sqrt{2h}})^4
\frac{i}{2\Delta^2\Delta_a \Delta_b }[(p_a \cdot p_a )\Delta_
b ^2\Delta_b ^{'}
+(p_b \cdot p_b )\Delta_a ^2\Delta_a ^{'}
+(p_a \cdot p_b )\Delta_a \Delta_b (\Delta_a ^{'}+\Delta_
b ^{'})] \times S. \nonumber \\
(c): & (\frac{\beta}{\sqrt{2h}})^4
\frac{i}{2\Delta^2\Delta_a \Delta_b^{'} }[(p_a \cdot p_a )(-\
\Delta_b {\Delta_b^{'}}^2)
-(p_b \cdot p_b )\Delta_a^{'} \Delta_a^2
+(p_a \cdot p_b )\Delta_a \Delta_b^{'}
(\Delta_a ^{'}+\Delta_b)] \times S. \nonumber
\end{eqnarray}
Again $S$ is a symmetry factor. In obtaining these formulae we remark that
some simplifications may arise. For example, in some cases
$q_i$ and $q_j$ happen to be the same, so reducing the the number
of integrations to be done. This is the case for the crossed box diagrams
(b) and (c) in figure 1, where $q_A=q_C$ or $q_B=q_D$, respectively.
Finally it is worth observing
that the simple pole residue of a crossed box diagram may
not vanish even though the corresponding
double pole residue vanishes.

\section*{IV. Simple pole residue at the double pole position}
It is known\cite{BCDSe}
that the double pole residue of the conjectured S-matrices
(\ref{SMAT}) is in agreement with perturbation theory assuming
$B(\beta)={\beta^2\over{2\pi}}+O(\beta^4)$.
Using the results of the previous section we will now
compare the nonleading simple poles. The following argument
shows that, at least to order $\beta^4$, the residue of a
simple pole of the exact
S-matrix at the general double pole position
vanishes and so we are left with showing the various contributions
from perturbation theory sum to zero.

The conjectured S-matrices for the ATFT are built in terms of
the building blocks $\{x\}$ (\ref{BBLOCK}). For $\theta$ away from
a pole and $B\ll1$ these blocks have an expansion
$\{x\}\sim 1- {\pi B\over 2h}
\sum_{\alpha ,\beta \in \lbrace 0,1\rbrace }(-1)\sp {\alpha +\beta}
\cot({-i\theta\over2}+
[(-1)\sp \alpha x +(-1)\sp \beta]{\pi\over  2h})$ while
near the pole $\theta_0={\pi\over h}[x\pm1]$ of $\{x\}$
we have
$\{x\}\sim [1\pm\frac{ i\pi B}{h}\frac{1}{(\theta-i\theta_0)}]
(1- {\pi B\over 2h}
\sum_{\alpha ,\beta \in \lbrace 0,1\rbrace }'
(-1)\sp {\alpha +\beta}\cot({-i\theta\over2}+
[(-1)\sp \alpha x +(-1)\sp \beta]{\pi\over  2h}))$,
where the prime means the singular term of the sum is not to be included.
Putting this together means that near  a double pole $i\theta_0$ the
S-matrices (\ref{SMAT}) have an expansion
\begin{equation}
S_{ab}= (\frac{\pi B}{h})^2\frac{1}{(\theta-i\theta_0)^2} \times
\bigl\{ c_0 + c_1(\theta-i\theta_0)B+\ldots\bigr\}, \quad c_0 = 1 + O(B).
\end{equation}
Here $c_1$ is a sum of cotangents whose precise form is not important
for the present argument.
All we must observe is that the residue of the simple pole
is proportional to $B(\beta)^3$. Using again no more than the tree
level result
$B(\beta)={\beta^2\over{2\pi}}+O(\beta^4)$ this means that
this simple pole vanishes to order $\beta^4$.

It now remains to be shown that the order $\beta^4$ contribution
to the simple pole residue of the S-matrix vanishes in
ordinary perturbation theory.
We  prove this using the Landau singularity analysis of the subleading
singularities already presented, together with the
leading singularity analysis of the box and triangle diagrams.
There are four kinds of contributions.
\begin{equation}
\begin{array}{l}
\mbox {(a) Feynman diagrams involving singular box diagram.} \nonumber \\
\mbox {(b) Feynman diagrams involving singular triangle diagrams.} \nonumber \\
\mbox {(c) Contributions from the expansion of Eq.(\ref{FLUX}).} \nonumber \\
\mbox {(d) Contributions from the expansion of Eq.(\ref{MAND}).} \nonumber
\end{array}
\end{equation}
We classify the contributing singular box diagrams in figure 1
with participating particles as follows ($k^*$ denotes the antiparticle of
$k$).

\begin{center}
\begin{tabular}{ccccc}
       & A   & B     & C   & D \\
(i)    & $a-k$ & $a+b-k$ & $b-k$ & $k^*$ \\
(ii)    & $a-k$ & $a+b-k$ & $a-k$ & $k^*$ \\
(iii)    & $a-k$ & $k^*$ & $b-k$ & $k^*$
\end{tabular}
\end{center}

Adding the three contributions from the three box diagrams in
Eq.(\ref{SLFBb}) and taking
into account the next order term in  the Jacobian given by Eq.(\ref{SLFBa}), we
get
\begin{equation}
i(\frac{\beta}{\sqrt{2h}})^4\frac{1}{(\theta-i\theta_0)}\cot\theta_0
+i(\frac{\beta}{\sqrt{2h}})^4\frac{1}{(\theta-i\theta_0)} \cot\theta_0
=i(\frac{\beta}{\sqrt{2h}})^4\frac{1}{(\theta-i\theta_0)} \cot\theta_0
\times 2. \nonumber
\label{A}
\end{equation}

The possible singular triangle diagrams are enumerated in figure 2.
Using the formula for the leading singularity of the triangle diagrams
given in \cite{BCDSe} together with the relevant vertex factors
(\ref{VER}) we easily obtain as the residues of the simple pole
\begin{eqnarray}
T1+T2 &=& i (\frac{\beta}{\sqrt{2h}})^4 \frac{1}{\sin\theta_0} \frac{-s_0}
{m_a m_b}, \\
T3+T4 &=& i (\frac{\beta}{\sqrt{2h}})^4 \frac{1}{\sin\theta_0} \frac{t_0}
{m_a m_b}.
\end{eqnarray}
Invoking the relation between the Mandelstam variables
\begin{equation}
s_0-t_0=4m_a m_b \cos\theta_0,
\end{equation}
we get the sum of the simple pole residues from the singular triangle diagrams
\begin{equation}
T1+T2+T3+T4= -i (\frac{\beta}{\sqrt{2h}})^4 \cot\theta_0 \times 4.
\label{B}
\end{equation}

Finally the contributions from
(c) and (d) are identical, each being given by
\begin{equation}
i (\frac{\beta}{\sqrt{2h}})^4 \cot\theta_0 \times 1.
\label{CD}
\end{equation}
Adding the four contributions in Eq.(\ref{A},\ref{B},\ref{CD})
gives the desired vanishing of the simple pole residue at the general
double pole positions,
\begin{equation}
i (\frac{\beta}{\sqrt{2h}})^4 \cot\theta_0\times(2-4+1+1)= ~0.
\end{equation}

\section*{V. Conclusions and Discussions}
We have in the ATFT a quantum field theory both rich in
structure and yet offering the tantalising possibility that it
may indeed be solved. The existence of quantum higher spin currents\cite{SY}
leads to the two-particle factorisation of the theories S-matrices.
Together with unitarity, analyticity and crossing,
the bootstrap equations of Zamolodchikov have been applied to these
theories to produce S-matrices. This bootstrap is still mysterious:
we do not know, for example, whether this is a genuine assumption
or may be proven within the axioms of field theory.
Nonetheless, the bootstrap appears to encode nonperturbative
information and the resulting S-matrices may be checked within
the context of standard perturbation theory.
Thus far they have passed every test applied.
This paper has provided a new and nontrivial test. These
theories have a complicated Landau singularity
structure. By deriving general expressions for the
subleading singularities of the box diagrams we have been able to compare
simple pole residues (ordinarily masked behind double poles)
from both field theory and the putative S-matrices.
For the $A_n$ theories dealt with here for simplicity
we find complete agreement.
The techniques developed are applicable to the wider class of
bosonic and supersymmetric\cite{DGZs}
ATFT. The outstanding question remains how the bootstrap
encodes the intricate cancellations and structure of the
renormalised field theory.

\section*{Acknowledgements}
This research is supported in part by KOSEF and Basic Science Research
Institute, Kyung Hee University.

\newpage
\section*{Figure and Table Captions}
Figure 1. Singular box diagrams and their dual diagrams.  \\
Figure 2. Diagrams involving singular triangle subdiagrams.

\newpage
\begin{picture}(150,110)(90,-65)
\put(130,-20){\line(0,1){40}}
\put(130,-20){\line(1,0){40}}
\put(170,20){\line(-1,0){40}}
\put(170,20){\line(0,-1){40}}
\put(150,20){\vector(-1,0){1}}
\put(130,0){\vector(0,-1){1}}
\put(150,-20){\vector(1,0){1}}
\put(170,0){\vector(0,1){1}}
\put(130,20){\line(-2,1){20}}
\put(120,25){\vector(2,-1){1}}
\put(130,-20){\line(-2,-1){20}}
\put(120,-25){\vector(2,1){1}}
\put(170,20){\line(2,1){20}}
\put(180,25){\vector(2,1){1}}
\put(170,-20){\line(2,-1){20}}
\put(180,-25){\vector(2,-1){1}}
\put(118,0){A}
\put(150,-32){B}
\put(175,0){C}
\put(150,24){D}
\put(120,30){a}
\put(120,-35){b}
\put(192,30){b}
\put(192,-35){a}
\put(145,-65){(a)}
\end{picture}
\begin{picture}(150,110)(90,-65)
\put(130,-20){\line(0,1){40}}
\put(130,-20){\line(1,1){18}}
\put(170,20){\line(-1,-1){18}}
\put(130,20){\line(1,-1){40}}
\put(170,20){\line(0,-1){40}}
\put(160,-10){\vector(-1,1){1}}
\put(130,0){\vector(0,-1){1}}
\put(160,10){\vector(1,1){1}}
\put(170,0){\vector(0,-1){1}}
\put(130,20){\line(-2,1){20}}
\put(120,25){\vector(2,-1){1}}
\put(130,-20){\line(-2,-1){20}}
\put(120,-25){\vector(2,1){1}}
\put(170,20){\line(2,1){20}}
\put(180,25){\vector(2,1){1}}
\put(170,-20){\line(2,-1){20}}
\put(180,-25){\vector(2,-1){1}}
\put(118,0){A}
\put(142,-17){B}
\put(175,0){C}
\put(138,14){D}
\put(120,30){a}
\put(120,-35){b}
\put(192,30){b}
\put(192,-35){a}
\put(145,-65){(b)}
\end{picture}
\begin{picture}(150,110)(90,-65)
\put(130,-20){\line(1,0){40}}
\put(130,-20){\line(1,1){18}}
\put(170,20){\line(-1,-1){18}}
\put(130,20){\line(1,-1){40}}
\put(170,20){\line(-1,0){40}}
\put(160,-10){\vector(1,-1){1}}
\put(150,20){\vector(-1,0){1}}
\put(160,10){\vector(1,1){1}}
\put(150,-20){\vector(-1,0){1}}
\put(130,20){\line(-2,1){20}}
\put(120,25){\vector(2,-1){1}}
\put(130,-20){\line(-2,-1){20}}
\put(120,-25){\vector(2,1){1}}
\put(170,20){\line(2,1){20}}
\put(180,25){\vector(2,1){1}}
\put(170,-20){\line(2,-1){20}}
\put(180,-25){\vector(2,-1){1}}
\put(141,11){A}
\put(150,-32){B}
\put(142,-17){C}
\put(150,24){D}
\put(120,30){a}
\put(120,-35){b}
\put(192,30){b}
\put(192,-35){a}
\put(145,-65){(c)}
\end{picture}

\begin{picture}(150,110)(90,-65)
\put(130,-30){\line(1,0){60}}
\put(160,-30){\vector(1,0){1}}
\put(160,-40){a}
\put(145,-20){$\Delta_a^{'}$}
\put(190,-30){\line(-1,3){20}}
\put(160,4){$\Delta_b^{'}$}
\put(170,30){\line(-1,0){60}}
\put(140,20){$\Delta_{a}$}
\put(110,30){\line(1,-3){20}}
\put(127,0){$\Delta_{b}$}
\put(120,0){\vector(-1,3){1}}
\put(110,-5){b}
\put(170,30){\line(-1,-1){20}}
\put(160,20){\vector(-1,-1){1}}
\put(110,30){\line(2,-1){40}}
\put(130,20){\vector(2,-1){1}}
\put(130,-30){\line(1,2){20}}
\put(140,-10){\vector(1,2){1}}
\put(190,-30){\line(-1,1){40}}
\put(170,-10){\vector(-1,1){1}}
\put(145,-65){(a')}
\end{picture}
\begin{picture}(150,110)(90,-65)
\put(130,-30){\line(1,0){60}}
\put(160,-30){\vector(1,0){1}}
\put(160,-40){a}
\put(190,-30){\line(-1,3){20}}
\put(170,30){\line(-1,0){60}}
\put(110,30){\line(1,-3){20}}
\put(120,0){\vector(-1,3){1}}
\put(110,-5){b}
\put(170,30){\line(-1,-1){20}}
\put(160,20){\vector(-1,-1){1}}
\put(110,30){\line(2,-1){40}}
\put(130,20){\vector(2,-1){1}}
\put(130,-30){\line(1,2){20}}
\put(140,-10){\vector(1,2){1}}
\put(190,-30){\line(-2,1){40}}
\put(170,-20){\vector(2,-1){1}}
\put(150,-10){\line(1,2){20}}
\put(160,10){\vector(1,2){1}}
\put(150,-10){\line(-1,-1){20}}
\put(140,-20){\vector(-1,-1){1}}
\put(145,-65){(b')}
\end{picture}
\begin{picture}(150,110)(90,-65)
\put(130,-30){\line(1,0){60}}
\put(160,-30){\vector(1,0){1}}
\put(160,-40){a}
\put(190,-30){\line(-1,3){20}}
\put(170,30){\line(-1,0){60}}
\put(110,30){\line(1,-3){20}}
\put(120,0){\vector(-1,3){1}}
\put(110,-5){b}
\put(170,30){\line(-1,-1){20}}
\put(160,20){\vector(-1,-1){1}}
\put(110,30){\line(2,-1){40}}
\put(130,20){\vector(2,-1){1}}
\put(130,-30){\line(1,1){20}}
\put(140,-20){\vector(-1,-1){1}}
\put(150,-10){\line(-1,1){40}}
\put(130,10){\vector(-1,1){1}}
\put(150,-10){\line(2,-1){40}}
\put(170,-20){\vector(2,-1){1}}
\put(190,-30){\line(-1,1){40}}
\put(170,-10){\vector(-1,1){1}}
\put(145,-65){(c')}
\end{picture}

\begin{center}
Figure 1. Singular box diagrams and their dual diagrams.
\end{center}
\vspace{0.5cm}

\begin{picture}(150,110)(20,-65)
\put(130,-20){\line(3,2){30}}
\put(145,-10){\vector(3,2){1}}
\put(140,-25){a+b-k}
\put(160,0){\line(-3,2){30}}
\put(145,10){\vector(3,-2){1}}
\put(140,20){k}
\put(130,20){\line(0,-1){40}}
\put(130,0){\vector(0,-1){1}}
\put(110,0){a-k}
\put(130,20){\line(-3,1){30}}
\put(115,25){\vector(3,-1){1}}
\put(90,25){a}
\put(130,-20){\line(-3,-1){30}}
\put(115,-25){\vector(3,1){1}}
\put(90,-30){b}
\put(160,0){\line(3,1){30}}
\put(175,5){\vector(3,1){1}}
\put(180,10){b}
\put(160,0){\line(3,-1){30}}
\put(175,-5){\vector(3,-1){1}}
\put(180,-15){a}
\put(140,-50){T1}
\end{picture}
\begin{picture}(150,110)(0,-65)
\put(130,-20){\line(3,2){30}}
\put(145,-10){\vector(3,2){1}}
\put(140,-25){a+b-k}
\put(160,0){\line(-3,2){30}}
\put(145,10){\vector(3,-2){1}}
\put(140,20){k}
\put(130,20){\line(0,-1){40}}
\put(130,0){\vector(0,-1){1}}
\put(110,0){a-k}
\put(130,20){\line(-3,1){30}}
\put(115,25){\vector(3,-1){1}}
\put(90,25){a}
\put(130,-20){\line(-3,-1){30}}
\put(115,-25){\vector(3,1){1}}
\put(90,-30){b}
\put(160,0){\line(1,0){20}}
\put(170,0){\vector(1,0){1}}
\put(162,-10){a+b}
\put(180,0){\line(3,1){30}}
\put(195,5){\vector(3,1){1}}
\put(200,10){b}
\put(180,0){\line(3,-1){30}}
\put(195,-5){\vector(3,-1){1}}
\put(200,-15){a}
\put(140,-50){T2}
\end{picture}

\begin{picture}(150,110)(20,-55)
\put(130,20){\line(2,-3){20}}
\put(140,5){\vector(2,-3){1}}
\put(120,0){a-k}
\put(150,-10){\line(2,3){20}}
\put(160,5){\vector(2,3){1}}
\put(165,0){b-k}
\put(170,20){\line(-1,0){40}}
\put(150,20){\vector(1,0){1}}
\put(150,25){k}
\put(150,-10){\line(-3,-1){30}}
\put(135,-15){\vector(3,1){1}}
\put(110,-20){b}
\put(150,-10){\line(3,-1){30}}
\put(165,-15){\vector(3,-1){1}}
\put(185,-20){a}
\put(130,20){\line(-3,1){30}}
\put(115,25){\vector(3,-1){1}}
\put(100,32){a}
\put(170,20){\line(3,1){30}}
\put(185,25){\vector(3,1){1}}
\put(200,25){b}
\put(140,-60){T3}
\end{picture}
\begin{picture}(150,110)(0,-55)
\put(130,20){\line(2,-3){20}}
\put(140,5){\vector(2,-3){1}}
\put(120,0){a-k}
\put(150,-10){\line(2,3){20}}
\put(160,5){\vector(2,3){1}}
\put(165,0){b-k}
\put(170,20){\line(-1,0){40}}
\put(150,20){\vector(1,0){1}}
\put(150,25){k}
\put(150,-10){\line(0,-1){20}}
\put(150,-20){\vector(0,1){1}}
\put(155,-20){b-a}
\put(150,-30){\line(-3,-1){30}}
\put(135,-35){\vector(3,1){1}}
\put(115,-55){b}
\put(150,-30){\line(3,-1){30}}
\put(165,-35){\vector(3,-1){1}}
\put(185,-55){a}
\put(130,20){\line(-3,1){30}}
\put(115,25){\vector(3,-1){1}}
\put(100,32){a}
\put(170,20){\line(3,1){30}}
\put(185,25){\vector(3,1){1}}
\put(200,25){b}
\put(140,-60){T4}
\end{picture}

\begin{center}
Figure 2. Diagrams involving singular triangle subdiagrams.
\end{center}

\end{document}